\documentclass[fleqn,usenatbib]{mnras}

\usepackage{newtxtext,newtxmath}

\usepackage[T1]{fontenc}
\usepackage{ae,aecompl}

\usepackage{graphicx}	
\usepackage{amsmath}	
\usepackage{amssymb}	
\usepackage{color}
\usepackage{multirow}
\usepackage{float}
\usepackage{adjustbox}
\usepackage{subfig}
\usepackage{float}
\usepackage{todonotes}

\defcitealias{Nealon:2018ic}{NDAMN18}

\title[Shadows in protoplanetary discs]{Scattered light shadows in warped protoplanetary discs}

\author[R. Nealon et al.]{\parbox{\textwidth}{
Rebecca Nealon$^{1}$\thanks{E-mail: rebecca.nealon@leicester.ac.uk},
Christophe Pinte$^{2}$,
Richard Alexander$^{1}$,
Daniel Mentiplay$^{2}$
\\
and Giovanni Dipierro$^{1}$}\vspace{0.2cm}
\\
$^{1}$Department of Physics and Astronomy, University of Leicester, University Road, Leicester, LE1 7RH, UK\\
$^{2}$Monash Centre for Astrophysics (MoCA) and School of Physics and Astronomy, Monash University, Clayton, Vic 3800, Australia
}

\date{Accepted XXX. Received YYY; in original form ZZZ}

\pubyear{2015}

\begin{document}
\label{firstpage}
\pagerange{\pageref{firstpage}--\pageref{lastpage}}
\maketitle

\begin{abstract}
Three-dimensional hydrodynamic numerical simulations have demonstrated that the structure of a protoplanetary disc may be strongly affected by a planet orbiting in a plane that is misaligned to the disc. When the planet is able to open a gap, the disc is separated into an inner, precessing disc and an outer disc with a warp. In this work, we compute infrared scattered light images to investigate the observational consequences of such an arrangement. We find that an inner disc misaligned by a less than a degree to the outer disc is indeed able to cast a shadow at larger radii. In our simulations a planet of $\gtrsim6$M$_{\rm J}$ inclined by $\gtrsim2^{\circ}$ is enough to warp the disc and cast a shadow with a depth of $\gtrsim 10\%$ of the average flux at that radius. We also demonstrate that warp in the outer disc can cause a variation in the azimuthal brightness profile at large radii. Importantly, this latter effect is a function of the distance from the star and is most prominent in the outer disc. We apply our model to the TW Hya system, where a misaligned, precessing inner disc has been invoked to explain an recently observed shadow in the outer disc. Consideration of the observational constraints suggest that an inner disc precessing due to a misaligned planet is an unlikely explanation for the features found in TW Hya.\end{abstract}




\section{Introduction}
Evidence is mounting that misalignments and warps in protoplanetary discs are common. Strongly misaligned discs present the most compelling examples, where the inner part of the disc is misaligned to the outer disc. The inner disc casts unique shadows on the outer disc, constraining the relative inclination between the two \citep{Min:2017oc}. Relative inclinations have been measured from ~30$^{\circ}$ for DoAR 44 \citep{Casassus:2018te} to 45$^{\circ}$ for AA Tau \citep{Loomis:2017do}, 70$^{\circ}$ for HD~142527 \citep{Marino:2015rh}, 72$^{\circ}$ for HD~100453 \cite{Benisty:2017kq} and up to 80$^{\circ}$ for HD~100546 \citep{Walsh:2017ic}. Numerical simulations suggested that such an arrangement may be generated through gravitational interaction with an inclined binary. For example, a circumbinary disc may be broken into two discs and the inner can cast asymmetric shadows onto the outer \citep{facchini_2013,Facchini:2017of,Zhu:2018vf}. For the specific case of HD~143006, \citet{Benisty:2018ve} demonstrated that such a model reproduces most of the observed features. Using HD~142527 as a case study, \citet{Price:2018pf} suggested that spiral arms, a cavity and shadows observed in some transition discs could be explained with the presence of an inclined binary. Characteristic disc structures including warps may also be generated through the interaction with an inclined external perturber, as in RW Aurigae system \citep{Dai:2015yr}, MWC 758 \citep{Dong:2018oq} and as shown by \citet{Cuello:2018bd}. Additionally, strongly misaligned discs may also be formed during independent chaotic accretion episodes \citep{Bate:2018ls}. 

Small to moderate misalignments have been invoked at the innermost edge of the disc to explain observations of dipper stars, e.g. \citet{Ansdell:2016ov} and recently by \citet{Pinilla:2018gb}. These may be driven by stellar magnetic fields that are misaligned to the disc, as in AA Tau \citep{Donati:2010be,Bouvier:1999uh} and LkCa 15 \citep{Alencar:2018ja}. Further evidence of misaligned discs is motivated by TW Hya, where recent infrared (IR) observations identified a variation of $\sim20\%$ in the surface brightness of the outer disc that moved with a speed much faster than Keplerian at that radius \citep{Roberge:2005be,Debes:2017fk,Poteet:2018be}. Such an asymmetry is difficult to explain as TW Hya is considered to be an isolated, single star system \citep{Mamjek:2005vw,Gagne:2017ve}--- broadly ruling out interactions with external perturbers or the interaction of the disc with a misaligned binary. \citet{Debes:2017fk} suggested that the feature in TW Hya may instead be caused by a misaligned, inner disc that is casting a shadow on the outer disc. In this interpretation the motion of the inner disc is governed by a misaligned planet that causes the disc to be inclined and rigidly precess, casting a shadow that moves with the precession rate of the inner disc. This model is consistent with existing observations; a small warp or misalignment in the inner region has been suggested  \citep[$\gtrsim10$~au]{Qi:2004bw,Pontoppidan:2008ge,Hughes:2011gw} and a planetary companion may explain the inner disc cavity \citep{Calvet:2002vr,Andrews:2016bw}. Additionally, a small warp is consistent with CO kinematics in the inner part of the disc \citep{Rosenfeld:2012uf}.

Numerical simulations have confirmed that this arrangement may be possible, as the disc structure can be altered by the influence of an inclined planet \citep{Nealon:2018ic,Zhu:2018vf}. In the case that the planet is able to open a gap, the initial disc is separated at the orbit of the planet into an inner and outer disc. A warp forms across the orbit of the planet, where a misalignment develops between the inner and outer disc \citep{Xiang-Gruess:2013fg,Bitsch:2013hg,Nealon:2018ic}. \citet{Xiang-Gruess:2013fg} demonstrated that the inner (and, in their case outer) disc tilt toward the orbit of the inclined planet. This behaviour is confirmed by \citet{Bitsch:2013hg} who found that the inner disc can achieve an inclination greater than that of the planet orbit. Both of these works agree that the disc responds most rapidly when the planet is at moderate inclinations (at inclinations $\sim$ the aspect ratio), rather than larger inclinations. Recently, \citet{Nealon:2018ic} conducted similar simulations with a larger outer disc radius (50~au rather than $\sim$20~au). We showed that the larger outer disc moves less under the influence of the misaligned planet, leading to a larger relative misalignment between the inner and outer disc. In the case that protoplanetary discs are larger than $\sim$20~au (\citealt{Andrews:2007pr} and \citealt{Ansdell:2018vi}, but see \citealt{Cieza:2019bw} and \citealt{vanTerwisga:2018ks}), this suggests that the inclinations between the inner and outer disc found in the previous works represent underestimates.

Scattered light observations of protoplanetary discs are the most important diagnostic to investigate the structural properties of protoplanetary discs. The scattering emission from small dust grains at the disc surface layer dominates at short wavelengths, roughly from optical to near-infrared (NIR) part of the spectrum. Small dust grains carry most of the opacity at optical and near-infrared wavelengths, and therefore control the stellar absorption and the disc’s temperature \citep{Hartmann:2009qu}. Additionally, these dust grains are expected to be well coupled with the gas in the inner and intermediate disc regions and populate the full vertical extent of the gas disc. Therefore, such observations can be considered a fundamental probe of the disc vertical structure \citep[see the review of][]{PP5_book} and can be used to infer the disc inclination, the shape of the surface and detect features at the surface layer \citep[e.g. induced by gaps carved by planets,][]{Jang-Condell:2012to}. The observational consequences of a low inclination warp in the inner disc have been explored by \citet{Arzamasskiy:2017ic}, where they compared the IR scattered light profile in the cases of an aligned and misaligned planet. In their intensity maps, asymmetries due to the warp from the misaligned planet were only observed in the inner regions of the disc. However, the duration of their simulation is shorter than in other works \citep[e.g.][]{Marzari:2009of,Xiang-Gruess:2013fg,Bitsch:2013hg,Nealon:2018ic}, the gap opened by the planet is not fully formed and the planet is kept at on a fixed orbit. This scenario is distinct from that invoked from observations of TW Hya, where the inner disc must be cleanly separated from the outer disc such that it can precess differentially.

In this paper we measure the observational consequences in scattered light for a disc warped by an interaction with a misaligned planet. We use the radiative transfer code {\sc mcfost} \citep{Pinte:2006nw,Pinte:2009ye} to generate scattered light images from the simulations presented in \citet{Nealon:2018ic} (hereafter \citetalias{Nealon:2018ic}). In Section~\ref{section:method} we summarise these simulations and our use of {\sc mcfost}. In Section~\ref{section:scattered_light} we present the scattered light results and examine the azimuthal brightness profile in the intermediate and outer disc. In Section~\ref{section:TW_Hya} we consider the observed features of TW Hya in the context of our model. In Section~\ref{section:disc} we discuss these results in the context of the limitations of our simulations and conclude in Section~\ref{section:concs}.

\begin{figure*}
	\includegraphics[width=\textwidth]{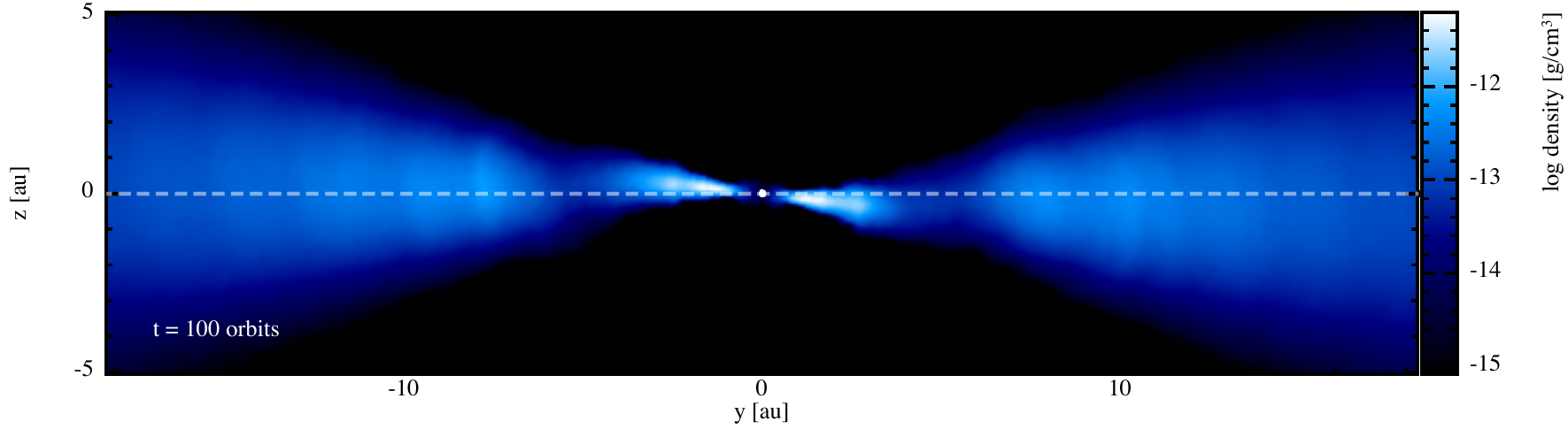}
	\caption{Density cross-section in the $x=0$ plane of our simulation from \citetalias{Nealon:2018ic} with a 6.5M$_{\rm J}$ planet inclined at $12.89^{\circ}$, located at $5$~au after $t=100$ orbits (the planet is not apparent in this figure as it is not located at $x=0$ at this time). The planet has separated the original disc into an inner and outer disc, both of which respond to the inclination of the planet. The inner disc tilts and precesses and the disc exterior to the planet orbit warps. The dashed line shows the position of the initial disc mid-plane.}
	\label{figure:density_cut}
\end{figure*}

\section{Numerical method}
\label{section:method}
\subsection{Three-dimensional hydrodynamic simulations}
In \citetalias{Nealon:2018ic} we conducted simulations using the smoothed particle hydrodynamics (SPH) code {\sc phantom}, a code particularly well suited to examining warps and disc-planet interactions \citep{Phantom}. The simulations from Section 4 of that work examined the behaviour of an accretion disc with a planet on a misaligned orbit. In the suite of simulations presented the properties of the disc were held constant; the disc extended from 0.1 to 50~au with a disc mass of $5 \times 10^{-3}$ M$_{\odot}$ and was simulated with $3.4 \times 10^6$ particles. The surface density profile in each case was initially set with a tapered inner edge and a power-law profile with an index of $p=-1$. The aspect ratio was set to $H/R=0.05(R/R_0)^{1/4}$ and a vertically isothermal disc assumed with the sound speed profile $c_{\rm s} = c_{\rm s,0} (R/R_0)^{-1/4}$, where $R$ is the cylindrical radius and $R_0$=1~au for the simulations we are considering. As the disc evolves, the temperature of the particles remains a function of radius and is assigned according to $c_s(R)$. The viscosity was modelled with the \citet{shakura_sunyaev} $\alpha$ parametrisation, with $\alpha = 10^{-3}$ for all the particles. We refer the interested reader to \citetalias{Nealon:2018ic} for further details.

A planet was placed in each disc at a distance of 5~au from the star with a mass of 0.13, 1.3 or 6.5M$_{\rm J}$ (corresponding to 0.1, 1.0 and 5.0$\times$ the thermal mass) and an orbital inclination of $2.15^{\circ}$, $4.30^{\circ}$ or  $12.89^{\circ}$ (0.5, 1.0 and 3.0$\times$ $H/R$, where $H$ is the scale height). This planet was then allowed to migrate and accrete mass due to the interaction with the surrounding disc for 200 orbits at the initial planet orbital radius. For those planets that were able to open a gap, we found \citep[in agreement with previous studies, see][]{Xiang-Gruess:2013fg} that the initial disc was separated into two discs. The inclination of the planets' orbit caused a change in the relative orientation of the inner and outer disc, with more massive and strongly inclined planets driving a larger difference.

Figure~\ref{figure:density_cut} shows the density cross-section $t=100$ planet orbits of our representative simulation (with a 6.5M$_{\rm J}$ planet inclined at $12.89^{\circ}$). Here the dashed line represents the initial position of the disc mid-plane. Although the outer disc (with $R>5$~au) has not moved appreciably, the inner has tilted towards the inclined planet orbit. This rotation is accompanied by the inner disc twisting as it undergoes differential precession. The outer disc also warps slightly in response to the misaligned planet (see Figures 8 and 9 of \citetalias{Nealon:2018ic}) and traces of the spiral shocks from the planet can be seen, e.g. around $y \sim -8$~au, $y\sim10$~au in the cross section in Figure~\ref{figure:density_cut}.

In our simulations, the most massive planet (6.5M$_{\rm J}$) was surrounded by a weak tidal stream that was most clearly visible during the initial phases where the gap is not fully carved. As this could potentially complicate any shadows due to the inclination of the inner disc, in this work we only consider snapshots where the planet is on the far side of the disc. We also supplement our parameter study with three new simulations. These have exactly the same properties as the previous set except that we consider a planet mass of 2.6M$_{\rm J}$ (corresponding to 2.0$\times$ the thermal mass). In line with our previous results the massive planet opens a gap and a warp between the inner and outer disc and driving precession of the inner disc.

For simplicity, in this work we use the $t\sim100$ orbit snapshots of the simulations presented in \citetalias{Nealon:2018ic}. At this time the gap opened by the planet is well established, as the relative tilt between the inner and outer disc is almost at maximum \citepalias[e.g. see Figure 8,][]{Nealon:2018ic}. The resolution of the simulations decreases at late times as the particles in the inner disc are accreted onto both the star and planet, but at $t\sim100$ orbits it has not decreased appreciably --- the shell averaged smoothing length to scale-height ratio $\langle h \rangle /H \approx 0.65$ at $R=3$~au and $< 0.2$ for $R > 10$~au. This choice means we do not take into account any temporal evolution of the discs, noting that the warp amplitude decreases slightly after $t=100$ orbits in all of our simulations.

\subsection{Radiative transfer calculations}
\label{subsection:mcfost_magic}
We use the continuum radiative transfer code {\sc mcfost} to generate scattered light images from our simulations \citep{Pinte:2006nw,Pinte:2009ye}. {\sc mcfost} is particularly well suited for use with  SPH simulations because it employs a Voronoi mesh rather than a cylindrical or spherical grid \citep{Camps:2015vj}. This method allows the mesh to be made without interpolation, as it is generated directly from the particle positions. To accommodate regions of the simulation that have lower numerical resolution, {\sc mcfost} implements a check based on the comparison between the size of the cell and the smoothing length of the particle. In the case that a dimension of any cell is three times larger than the smoothing length, {\sc mcfost} considers the region outside three smoothing lengths of the particle within that cell to be optically void.

We focus on scattered light observations that are sensitive to the distribution of micron sized dust grains. As these grains are very well coupled to the gas, we can use the gas distribution in our simulations directly. From our simulations we estimate a Stokes number of $\sim1.2\times10^{-3}$ in the upper layers for $\sim \mu$m grains at 5~au, confirming this assumption is valid. The total mass of dust in the disc is calculated from the mass of gas in the disc (directly measured from the simulation) and an assumed dust-to-gas ratio. We assume that each of the dust grains are spherical and homogeneous (according to Mie theory), that the dust and gas is in thermal equilibrium and the dust opacities are independent of temperature. We assume a dust-gas ratio of 100, where the dust grains have a power-law size distribution with an exponent of -3.5 \citep{Mathis:1977bw} between 0.03 and 1000$\mu$m using 100 grain sizes.

To improve efficiency the computation of scattered light is broken into two steps. First, the temperature of the disc is calculated for each of the Voronoi cells. Photon packets are emitted from the central star using a Monte Carlo approach and the path of each photon packet is governed by scattering, absorption and re-emission processes \citep[see][]{Pinte:2006nw}. Second, the scattered light is calculated by emitting photon packets randomly from the disc and central star. In this second step the photon packets are always scattered (i.e. there is no absorption) but the energy of the packet is multiplied by the albedo at each scattering event. Images are then produced with a ray-tracing method, where we integrate the radiative transfer equation using the specific intensities and temperatures computed by the Monte-Carlo runs. We refer to \citet{Pinte:2006nw} and \citet{Pinte:2009ye} for a more detailed description of {\sc mcfost}.

Unless otherwise stated, we use the following values for our {\sc mcfost} calculations. To compute the temperature structure and the scattered light we use $10^8$ photon packets and our grid is set by the Voronoi mesh. We set our image to have a size of 120~au and define the angular momentum vector of the outer disc as direction of the viewer (that is, the viewer is `face-on' to the outer disc). We do not include any contribution from viscous heating. The luminosity of the star is calculated from 1~Myr Siess isochrone \citep{Siess:2000vd}, assuming a fiducial 1M$_{\odot}$ star. This corresponds to an effective temperature of $T_* = 4278$K and a luminosity of $L_* = 0.16$L$_{\odot}$. The scattered light images have been computed at a wavelength of $\lambda = 1.6\mu$m and convolved with a 2D Gaussian to smooth across 1.8~au (which corresponds to a FWHM of 30mas at a distance of 60~pc, \citealt{Gaia_Collaboration:2018cr}). We chose these values as they represent the angular resolution of TW Hya using a typical instrument, e.g. SPHERE or GPI. The central 5~au of the disc is masked in figures (grey inner circle) to simulate the effect of a coronograph.

\subsection{Measuring the surface brightness as a function of angle}
\label{subsection:method_azimuthal_profile}
Using the NICMOS and STIS instruments on the Hubble Space Telescope (HST), \citet{Roberge:2005be}, \citet{Debes:2017fk} and \citet{Poteet:2018be} identified subtle fluctuations in the IR surface brightness of TW Hya by considering the flux as a function of position. In order to measure this, they separated the disc into concentric annuli (describing the distance from the star) as well as azimuthal wedges (describing the angle around the disc). This allows the flux in each radial annulus to be described as a function of the azimuthal angle. We follow the method outlined by \citet{Debes:2017fk} in order to describe the flux from our simulations in a comparable manner.

We discretise the outer disc into wedges where the annuli are delimited by 5, 15, 30 and 50~au and the azimuthal slices have an angular size of $5^{\circ}$. In each single wedge, the flux from all pixels is averaged and the standard deviation quantifies the noise in the flux at that location in radius and azimuthal angle. The flux across each radial annulus is averaged and we use it to normalise the flux in each wedge in that radial annulus. This means that the amplitude of the surface brightness profiles plotted as a function of azimuthal angle represent the variation from the average brightness in that radial annulus. Following the method of \citet{Debes:2017fk}, we assume this azimuthal profile can be fitted with a sinusoidal function of the form
\begin{align}
F(\phi) = A \cos \left[ \phi - C \right] + 1,
\label{equation:fit}
\end{align}
where $A$ and $C$ are unconstrained and $\phi$ is the angle measured from the positive $x$-axis in an anti-clockwise direction. This form is scaled to the average brightness around the annulus, where $A$ determines the magnitude of the variation compared to the average and the position angle $C$ is the angular location of the strongest variation.

The fit is achieved using a least-squares curve fitting tool {\sc scipy.optimize} \citep{Jones:2001aa}. When conducting the fit, each measurement of the flux at a specific azimuthal wedge is weighted by the standard deviation of the flux in that wedge. Each parameter ($A$ and $C$) is returned with an associated standard deviation. We use the standard deviation of the parameter $A$ as an estimate for our accuracy of the fit on the amplitude of the surface brightness profile.

\begin{figure}
	\includegraphics[width=\columnwidth]{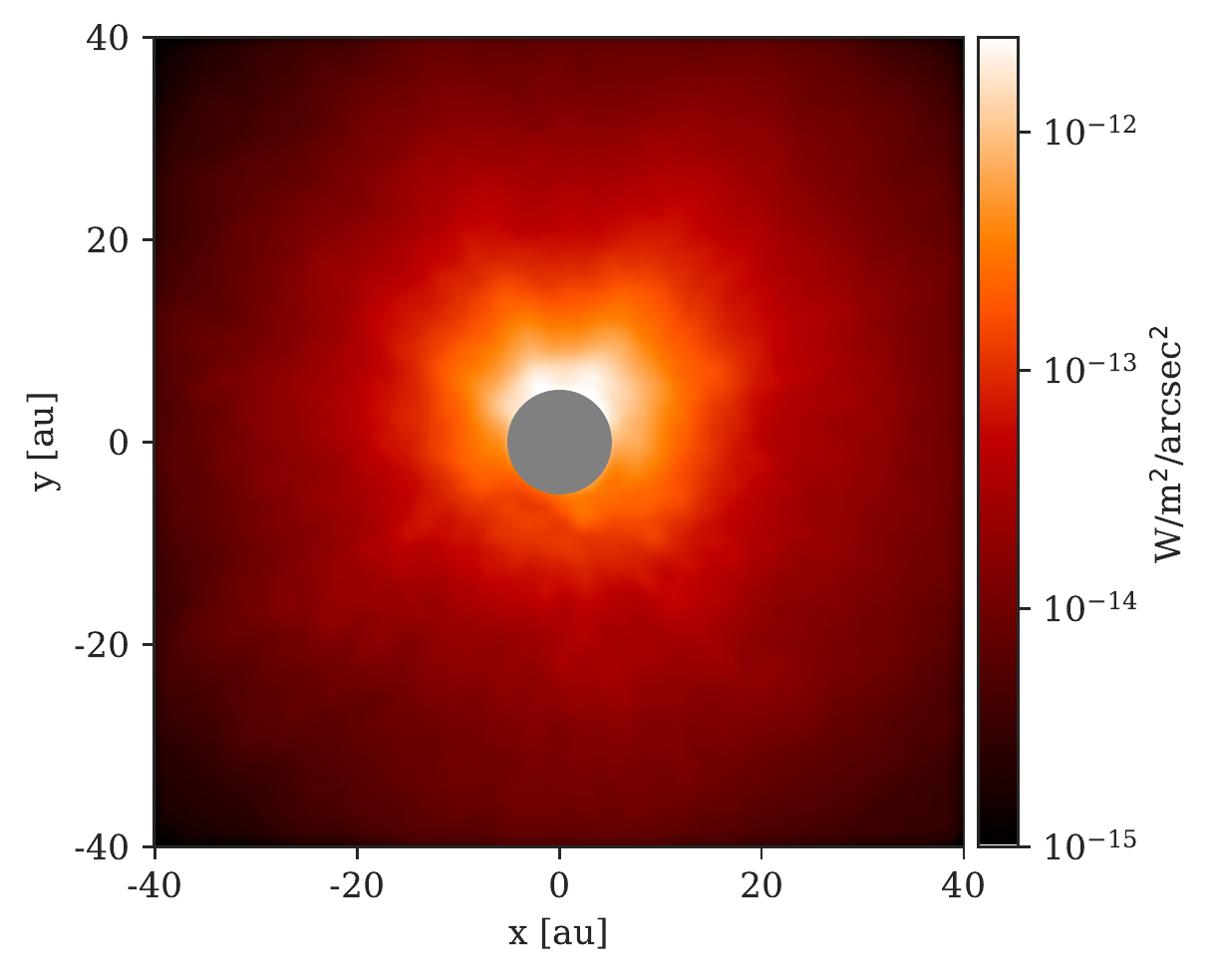}
   	\caption{Total intensity in scattered light at 1.6$\mu$m from a misaligned inner disc, calculated from the simulation shown in Figure~\ref{figure:density_cut}. The upper right side appears brighter, consistent with shadowing from the inner disc at $R<5$~au. Here the inner 5~au has been masked with a grey circle.} \label{figure:reference}
\end{figure}

\begin{figure*}
\hspace*{-0.5cm}
	\includegraphics[width=0.9\textwidth]{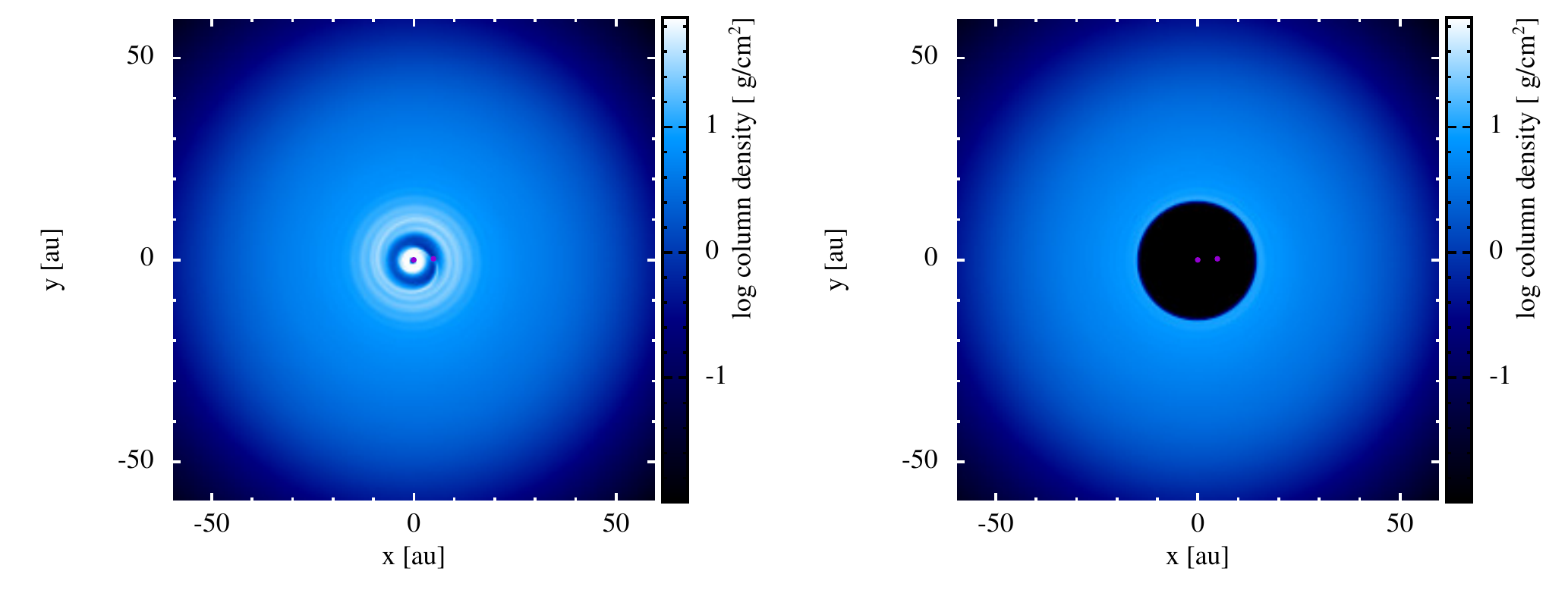}
\hspace*{0.7cm}
	\includegraphics[width=0.92\textwidth]{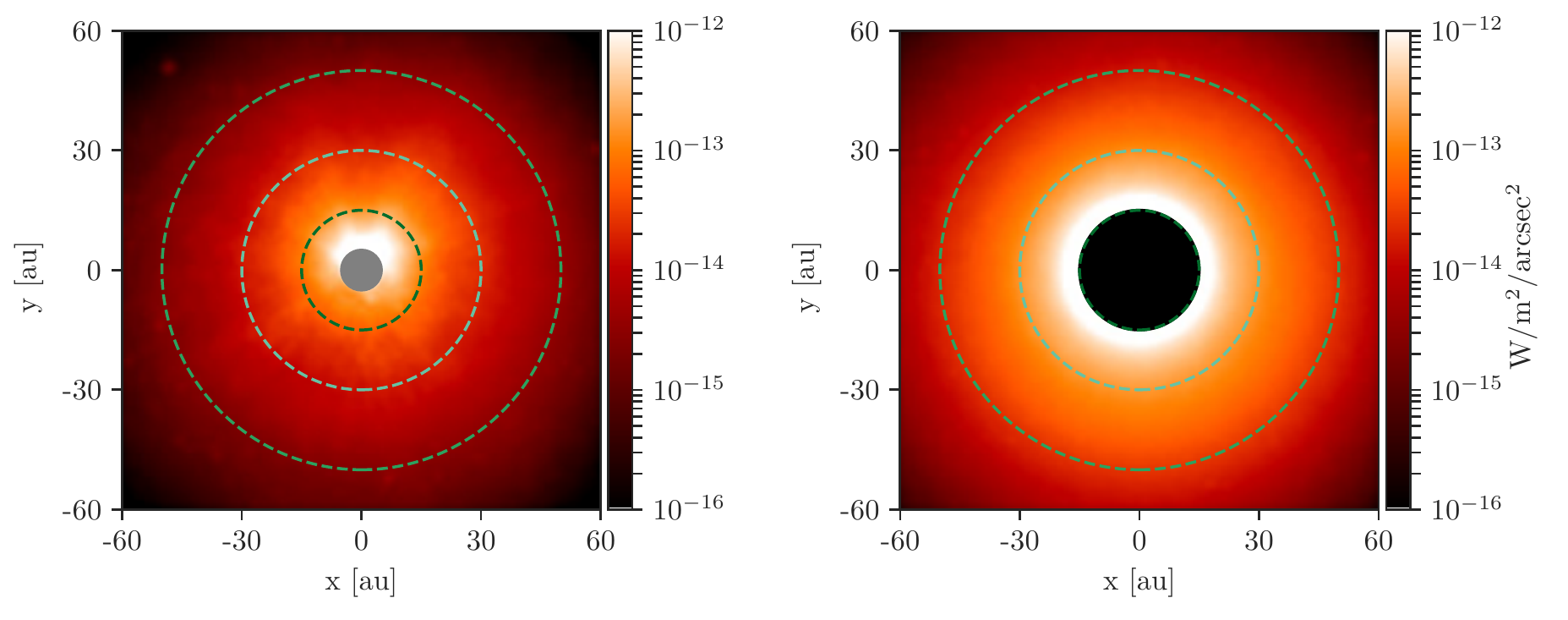}
\hspace*{-0.6cm}
	\includegraphics[width=0.9\textwidth]{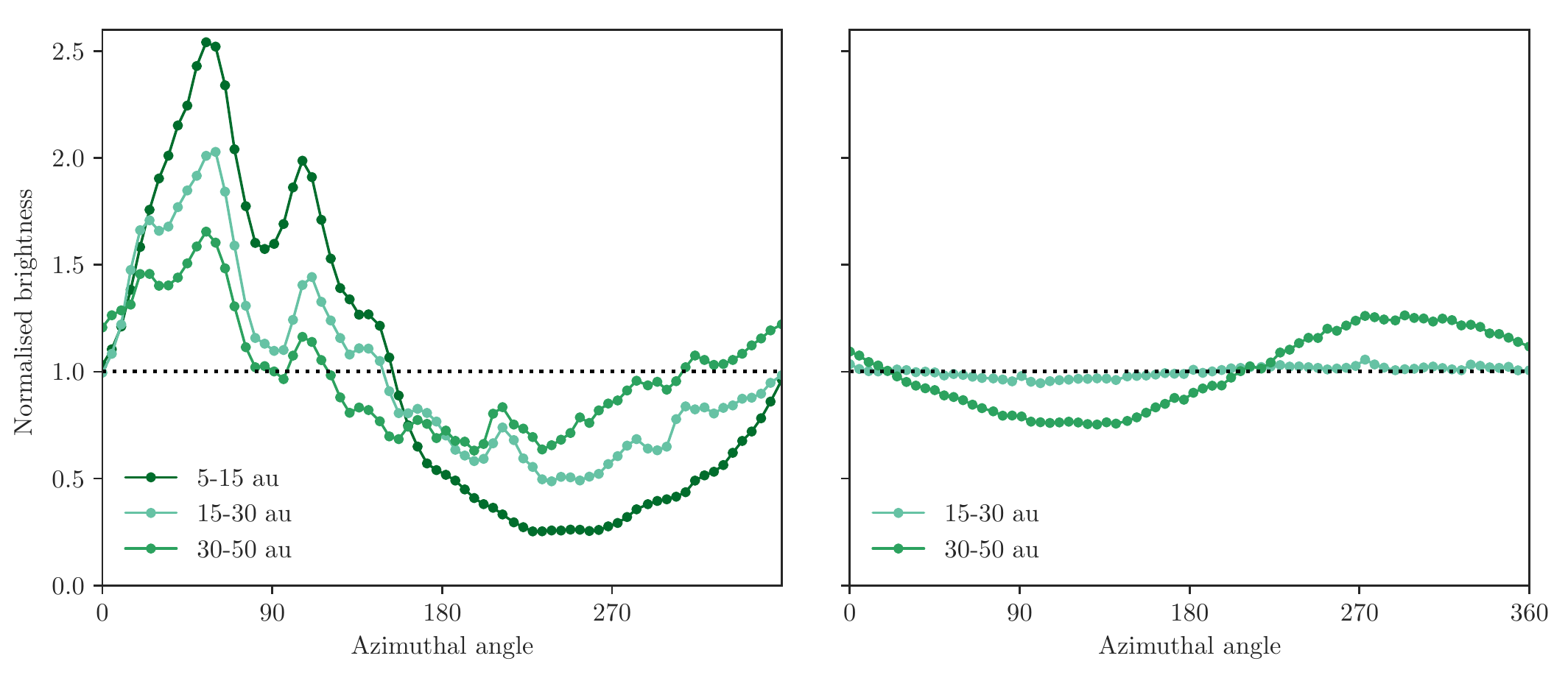}
	\caption{Casting shadows in warped discs: the upper panels show the column density of a simulation from \citetalias{Nealon:2018ic} containing a 6.5M$_{\rm J}$ planet inclined at $12.89^{\circ}$ at 5~au at $t=100$ orbits. The middle panels show the scattered light image recovered from this simulation, with concentric annuli indicated with dashed lines. The lower panels show the azimuthal surface density profiles from each of these annuli, normalised by the average brightness in the annulus. The left panels show the shadow cast from the misaligned inner disc. In the right panels, gas from $R<15$~au has been removed (as a post-processing step), demostrating variation in the surface density brightness due to the warp induced in the disc by the misaligned planet.}\label{figure:reference_composite}
\end{figure*}

\begin{figure*}
	\includegraphics[width=\textwidth]{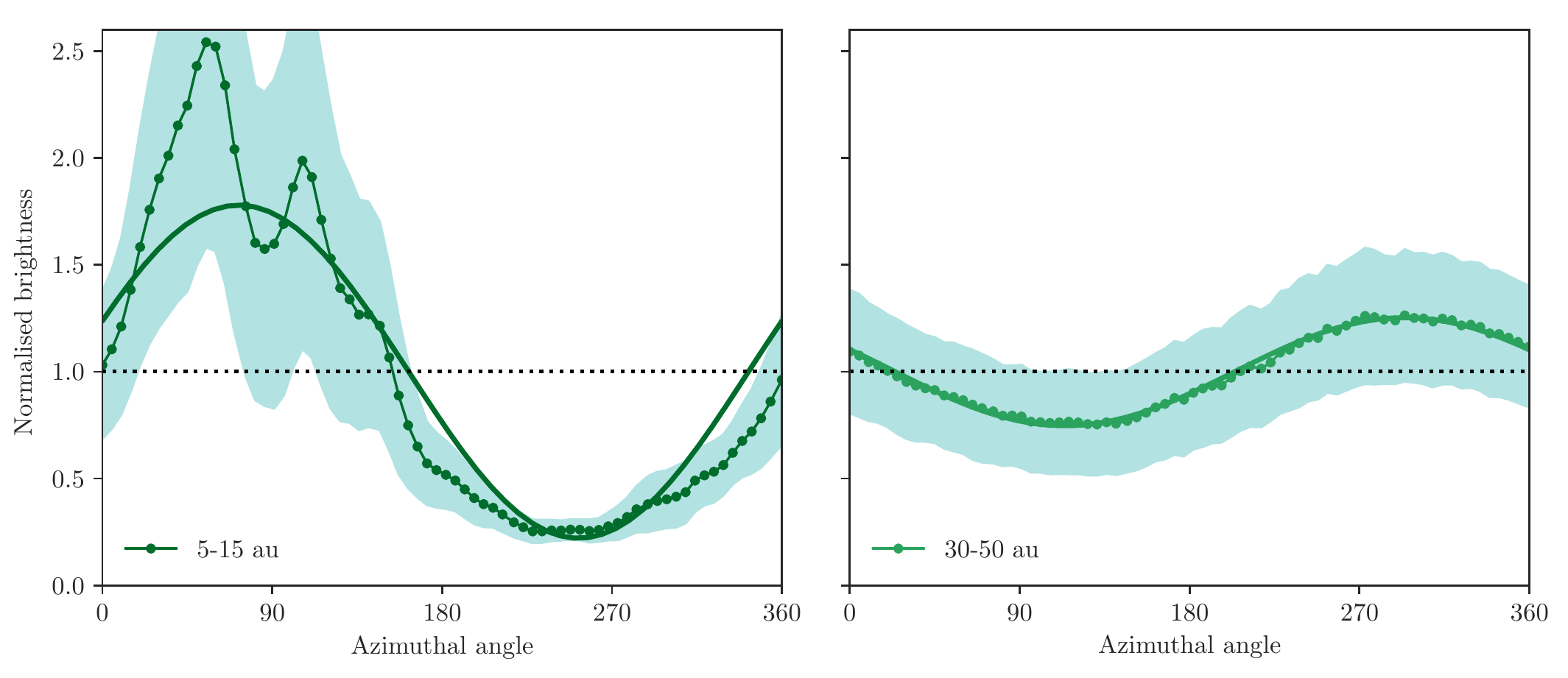}
   \caption{Measuring the amplitude of the surface brightness variation, with the thick solid lines representing the fit to the data from Equation~\ref{equation:fit}. The left panel shows the variation due to a misaligned inner disc (and warp) and the right panel the variation due to warping in the outer disc, both for the simulation shown in Figure~\ref{figure:reference_composite}.}\label{figure:fitting_example}
\end{figure*}

\section{Synthetic scattered light observations}
\label{section:scattered_light}
We computed the scattered light images for our 12 simulations described above using {\sc mcfost}. Figure~\ref{figure:reference} shows an example of the resulting total intensity image for the simulation with a 6.5M$_{\rm J}$ planet initially inclined at $12.89^{\circ}$. It is apparent that the upper, right side of this image is more strongly  illuminated than the lower, left side. This brightness feature is consistent with the orientation of the misaligned inner disc casting a shadow towards the left, lower side of the disc in Figure~\ref{figure:reference}. We examine the properties of this shadow in the following sections.

\subsection{Azimuthal brightness profile}
The left panels of Figure~\ref{figure:reference_composite} show the density rendering of our simulation (upper), the corresponding scattered light image (middle, as in Figure~\ref{figure:reference}) and the azimuthal brightness profile (lower). The surface brightness has been measured across 5-15~au, 15-30~au and 30-50~au as indicated on the scattered light image. The surface brightness profile shows asymmetries are present at all radii outside of the planet orbit, with the deepest shadow occurring near the inner disc. The profile of the shadow is quite noisy for $\phi < 180^{\circ}$, likely due to low resolution in the particle distribution close to the star being strongly illuminated.

In order to test the effect of the inner disc, we remove all the gas inside of 15~au as a post-processing step and re-run the radiative transfer calculation. The density rendering, scattered light image and surface brightness profile are shown in the right panels of Figure~\ref{figure:reference_composite}. We note that the same profile shown in Figure~\ref{figure:reference_composite} is achieved by only removing the gas inside of 5~au, but with 15~au we can be confident there are no effects due to warping at the inner edge of the outer disc. By eliminating all gas in the inner region, we remove any structures that could cast a shadow such that the outer disc is uniformly illuminated by the star. The surface brightness still shows significant but much smaller variations in the azimuthal profile (lower, right panel). This variation is due to the influence of the misaligned planet on the structure of the disc; the planet induces a warp in the outer disc and this causes one side of the disc to be illuminated more strongly. Because the illumination of the disc depends on the differential twist between the source of the illumination (e.g. the star) and the radius of the disc that is being illuminated, this affect appears stronger at larger radii. The lower, right panel of Figure~\ref{figure:reference_composite} demonstrates this, with the variation between 30-50~au stronger than in the intermediate disc. Due to warp propagation, we note that at later times in our simulation the warp amplitude is lower in the outer disc.

To confirm this variation in brightness in the outer disc is exclusively caused by differential twist in the disc driven by the planet, we first repeat the radiative transfer but force the scattering to be isotropic. Here we recover the same variation in the surface brightness, confirming that this effect is due to the preferential illumination of the disc. Second, we repeat this calculation at $t=40$ orbits when the inner disc is inclined but the outer disc has not yet been able to twist up \citepalias[see Figures 8 and 9,][]{Nealon:2018ic}. At this earlier time when the outer disc is not yet warped, we find no variation in the azimuthal brightness profile.

Our simulations with a misaligned planet therefore demonstrate that a non-axisymmetric azimuthal brightness profile can be driven by two mechanisms. First, the inner misaligned disc can cast strong shadows that emanate from the inner region. Second, the outer disc warp causes some part of the disc to be preferentially illuminated with the strongest variation occurring at large radii. Understanding the shadow in TW Hya requires us to consider both of these effects together.

\subsection{Amplitude of the brightness variation}
We quantify the variation in the surface brightness profile observed in Figure~\ref{figure:reference_composite} using the curve fitting procedure outlined in Section~\ref{subsection:method_azimuthal_profile}.  Figure~\ref{figure:reference_composite} suggests that a variation could be caused by either shadowing from the inner disc or warping of the outer disc, so we present fits for both of these. Figure~\ref{figure:fitting_example} shows the azimuthal brightness for both cases with the variation due to the misaligned inner disc in the left panel and the variation measured in the outer disc in the right panel. Both are shown with the standard deviation in the flux measurement as a shaded region. We measure the variation from the misaligned inner disc between 5-15~au, where there is little confusion from the outer disc warp and the shadow is strongest. We also measure the variation due to the outer disc warp between 30-50~au, where it is strongest (and after eliminating the inner disc, as described above). Both measurements thus represent an upper limit on the amplitude of the shadow for $t=100$ orbits and should be considered as demonstrations of the effect of the warp.

For a 6.5M$_{\rm J}$ planet inclined at $12.89^{\circ}$, the intensity of the shadow cast by the misaligned disc between 5-15~au is best fit with an amplitude of 78\%$\pm$3\% (compared to the mean in this annulus). For the same simulation, in the outer disc between 30-50~au we measure a variation in the surface brightness caused by the outer disc warp of 25\%$\pm$9\%. It is clear that the fit to the shadow in the left panel of Figure~\ref{figure:fitting_example} is not a good representation of the actual profile cast by the shadow. This is particularly apparent for azimuthal angles $\lesssim 120^{\circ}$, where the profile is dominated by noise likely due to the strong illumination of low resolution particles. However we chose a simple sinusoid function because of its simplicity and also to allow direct comparison to the observations in \citet{Debes:2017fk} and \citet{Poteet:2018be}, as they used this form.

\begin{figure*}
	\includegraphics[width=\textwidth]{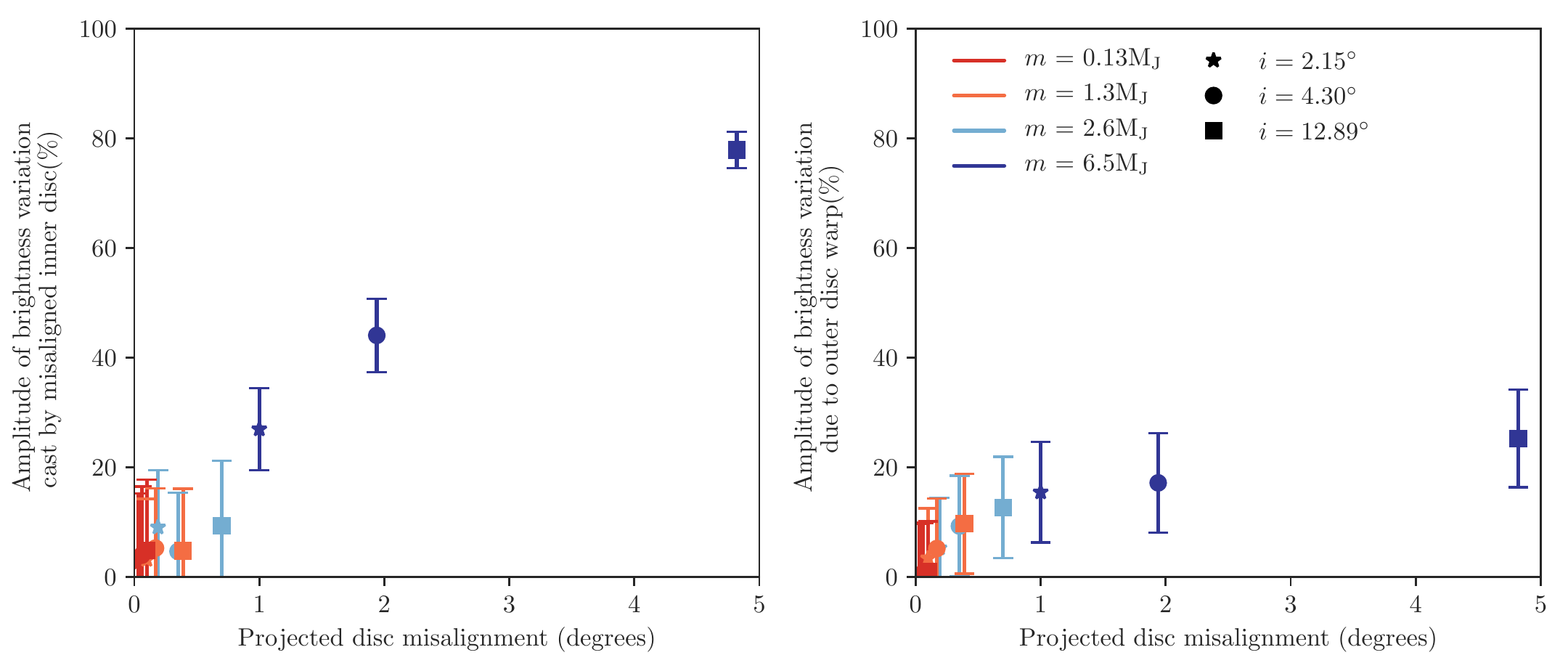}
   \caption{What planet mass and inclination leads to the largest shadow? The percentage is derived from the curve fitting procedure (i.e. the amplitude of the fitted curve in Figure~\ref{figure:fitting_example}) over our 12 simulations at $t\sim100$ orbits. The left panel shows the shadow cast by the inclined inner disc and the right panel the variation due to the warp in the outer disc. The different planet masses and inclinations are shown by the legend in the right hand panel.}\label{figure:parameter_sweep_outer_shadow}
\end{figure*}

We repeat this measurement of the variation in the surface brightness due to either the misaligned inner disc or warp in the outer region for each of our 12 simulations. Figure~\ref{figure:parameter_sweep_outer_shadow} shows the amplitudes from each of these fits with their associated uncertainties, plotted as a function of the relative inclination between the inner and outer disc (i.e. the maximum projected inclination, calculated as the angle between their average angular momentum vectors). The depth of the shadow driven by the inner disc (left panel, Figure~\ref{figure:parameter_sweep_outer_shadow}) is largest for more massive planets. This agrees with our previous findings where the massive planets are able to open a gap more cleanly, allowing a larger difference in the orientation between the inner and outer disc \citepalias{Nealon:2018ic}. For those planets not able to cleanly separate the inner and outer disc (in our simulations, planets with $0.13$M$_{\rm J}$), we find that the amplitude of the shadow depths measured are consistent with zero.

Figure~\ref{figure:parameter_sweep_outer_shadow} thus suggests that an appreciable shadow (i.e. more than $\sim$10\%) is only cast by planets much more massive than the gap-opening mass. For an aligned planet, the gap-opening mass is approximately the same as the thermal mass of 1.3M$_{\rm J}$ \citep{Crida:2006bv,PP6_book}. With the parameters used in our simulations, the limit for generating a brightness variation corresponds to $\gtrsim6$M$_{\rm J}$ planets (at inclinations $\gtrsim 2^{\circ}$) that drive misalignments of only a few degrees between the inner and outer disc. We find a similar limit exists on the amplitude of the variation driven by the warp in the outer disc (right panel, Figure~\ref{figure:parameter_sweep_outer_shadow}); planets that are more massive than the gap clearing mass are able to drive appreciable variations in the brightness profile in the outer disc. Additionally, in \citetalias{Nealon:2018ic} we showed that the warp created across the planet orbit for a massive planet at low/moderate inclinations was mainly caused by differential precession of the inner disc. These shadows therefore represent the misalignment due to the precession of the inner disc, rather than the inner disc being strictly tilted away from the outer disc.

\subsection{Effect of viewing inclination}
\begin{figure*}
	\includegraphics[width=\textwidth]{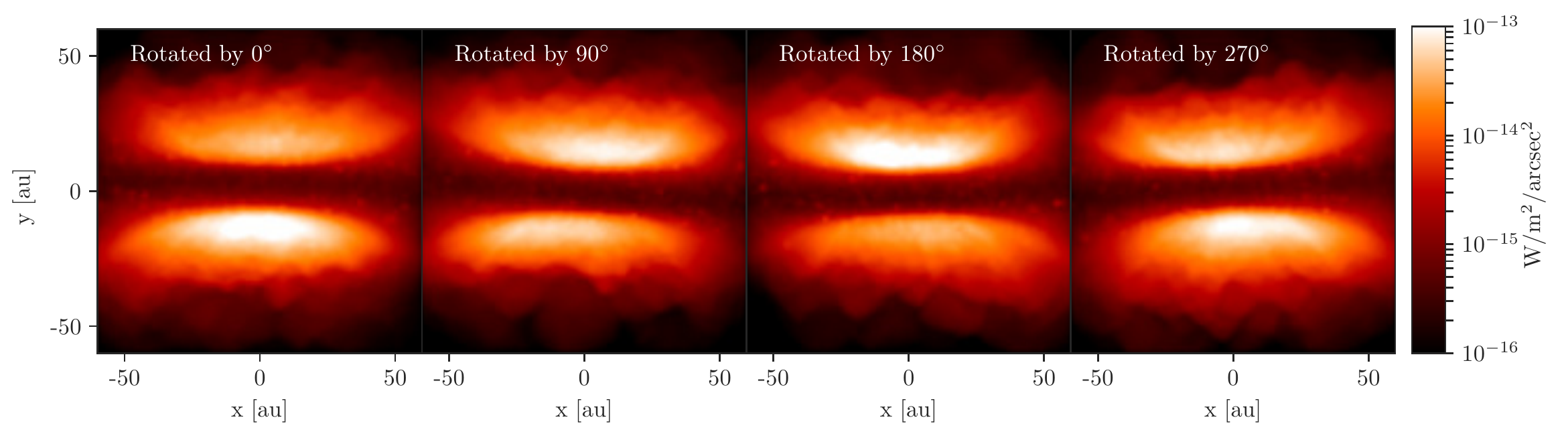}
   \caption{Scattered light surfaces when the disc in Figure~\ref{figure:reference} is viewed edge-on from different sides (with the rotation indicated measured clock-wise from Figure~\ref{figure:reference_composite}). The asymmetry in the flux between the upper and lower disc surfaces is mainly due to the warp in the outer disc.} \label{figure:inc_comparison}
\end{figure*}

In the above results we have assumed the viewing inclination is parallel to the angular momentum vector in the outer disc (i.e. the outer disc is face on). However, as the variation in surface brightness is a geometric effect it is likely to change when viewed at other inclinations. A cursory examination of the disc at different inclinations shows that this is indeed the case with shadows only clear for viewing inclinations less than about $30^{\circ}$. At inclinations larger than this, brightness profiles are complicated by the dust scattering function and flaring structure.

We thus consider the extreme case of viewing the disc edge-on (i.e. the viewing inclination is $i=90^{\circ}$ and the angular momentum vector of the outer disc is perpendicular to the line of sight). As our simulation does not go for a full precession of the inner disc, we emulate precession by rotating the disc through $90^{\circ}$, $180^{\circ}$ and $270^{\circ}$, measured clock-wise from the orientation in Figure~\ref{figure:reference_composite}. When the disc is subsequently turned edge on, the inner disc is then at four different orientations, as it would be across a full precession.

Figure~\ref{figure:inc_comparison} shows the inclined view of the entire disc. There is a flux asymmetry between the upper and lower disc surfaces depending on the rotation of the disc, with the lower disc surface brighter with an viewing rotation of $0^{\circ}$ and the upper disc surface at a rotation of $180^{\circ}$. To confirm whether this asymmetry is caused by the misaligned inner disc or the warp in the outer disc, we repeat this calculation but remove the gas inside of 15~au. As before, this removes any shadowing that could be caused by the inner misaligned disc such that any changes in flux should only be caused by the warp in the outer disc. We find that the disc is brighter in absolute flux (as the remaining disc is now more strongly illuminated) but that the same flux asymmetry pattern appears. This comparison suggests that a flux asymmetry between the upper and lower surfaces of an edge-on disc is a potential  signature of a disc warp. If the warp is steady, in observations these asymmetries would evolve according to the precession time-scale of the warp in the outer disc. We note that similar asymmetries have already been observed on much shorter time-scales, e.g. HH 30 \citep{Stapelfeldt:1999br}.

\subsection{Temperature structure}
\label{subsection:temp_structure}
\begin{figure}
	\includegraphics[width=\columnwidth]{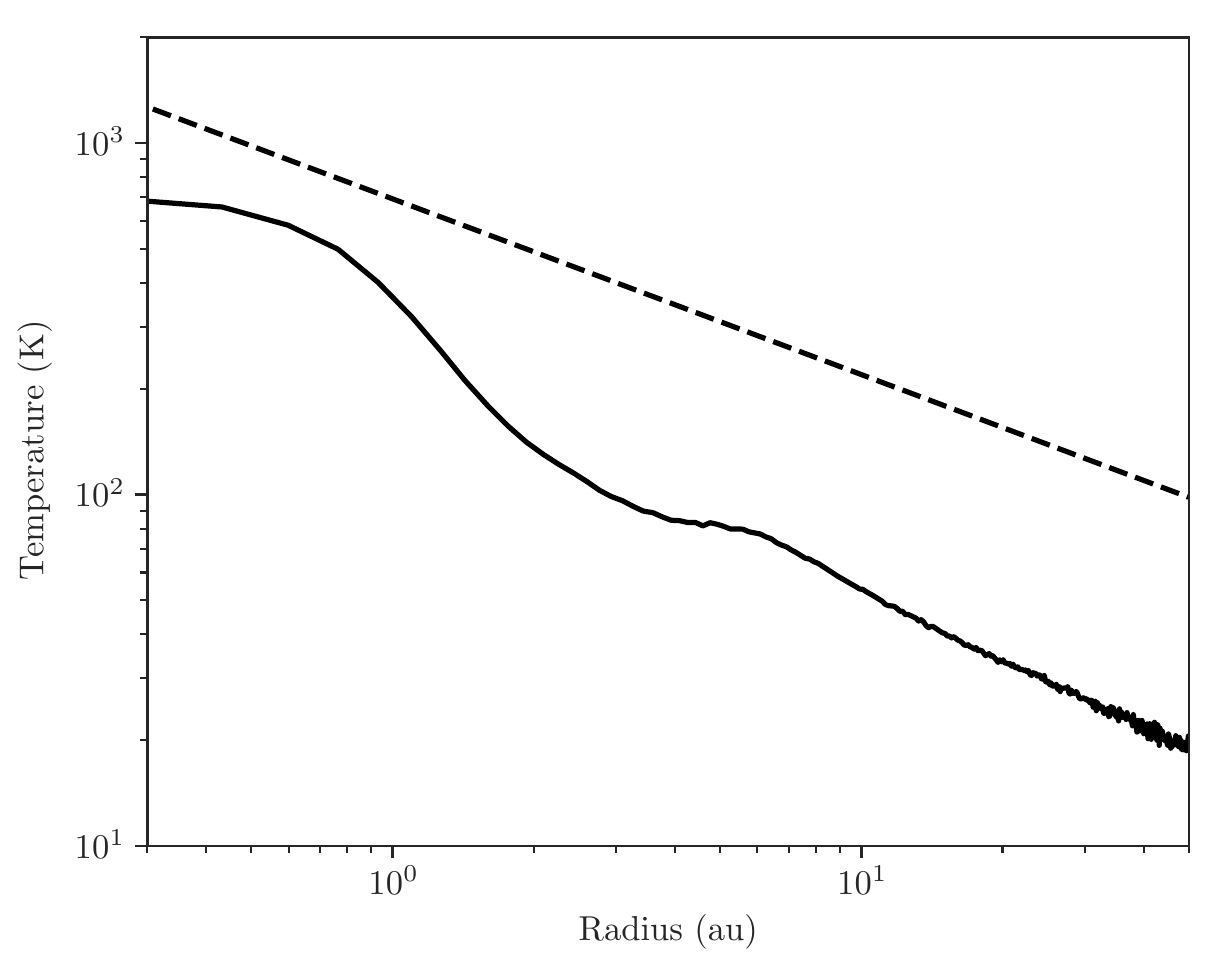}
	\caption{Comparison of the mid-plane temperature structure used to generate the scattered light images (solid line) with the vertically isothermal approximation assumed in \citetalias{Nealon:2018ic} (dashed line). Although the temperature profile calculated by {\sc mcfost} has the same radial power-law slope, it is cooler than that assumed in our simulations. The feature at 5~au corresponds to the planet orbit, with spiral waves emanating from this region.}
	\label{figure:midplane_temp}
\end{figure}

While our simulations in \citetalias{Nealon:2018ic} assume the disc is vertically isothermal, {\sc mcfost} calculates the temperature structure due to the illumination by the star. To quantify the effect of our assumption, Figure~\ref{figure:midplane_temp} compares the temperature calculated by {\sc mcfost} for each Voronoi cell in the disc mid-plane with the temperature predicted in our simulation. Here we have discretised the Voronoi mesh into annuli (defined by cylindrical radius) and by comparing with the azimuthally averaged profile of the angular momentum, extracted and averaged the temperature of only those cells that are within a distance of $|z|<0.05$ from the mid-plane. Thus the profile in Figure~\ref{figure:midplane_temp} represents the temperature of the (sometimes misaligned) disc mid-plane. Outside of the planet orbit (at $R \gtrsim 10$~au) the radial temperature gradient is consistent with that assumed in the hydrodynamical simulations (i.e. $T(R) \propto R^{-1/2}$). Around the planet orbit (at 5~au) the spiral density waves created by the planet are identified in the temperature structure. The disc inside the planet orbit and around the planet is warped and not uniformly illuminated by the star, hence we do not expect good agreement in this inner region.

The offset in Figure~\ref{figure:midplane_temp} demonstrates that the temperature assumed in the simulation is hotter than that calculated using {\sc mcfost}, by a factor of $\sim 4$. This offset corresponds to a difference in the sound speed in the disc of $\sim 2$ such that communication (via wave propagation) is occurring about twice as fast than it would if radiative transfer were included. A change in the sound speed (or equivalently, the thickness of the disc) may prevent the planet carving a gap or change the precession behaviour. However, for the parameters we have simulated in \citetalias{Nealon:2018ic} we estimate that such a small change does not affect the relationship between the sound crossing and precession time-scales. As this relationship determines the evolution of the disc (in the event the planet is massive enough to open a gap, whether the inner disc will precess with respect to the outer disc), we thus expect that a cooler disc will exhibit similar dynamical evolution. Additionally, the disc temperature calculated by {\sc mcfost} is primarily determined by the luminosity of the star, which in turn depends on the input parameters of its age and mass. A higher luminosity star would reduce the discrepancy (our adopted value is $L_*=0.16$~L$_{\odot}$).

\begin{figure}
	\includegraphics[width=\columnwidth]{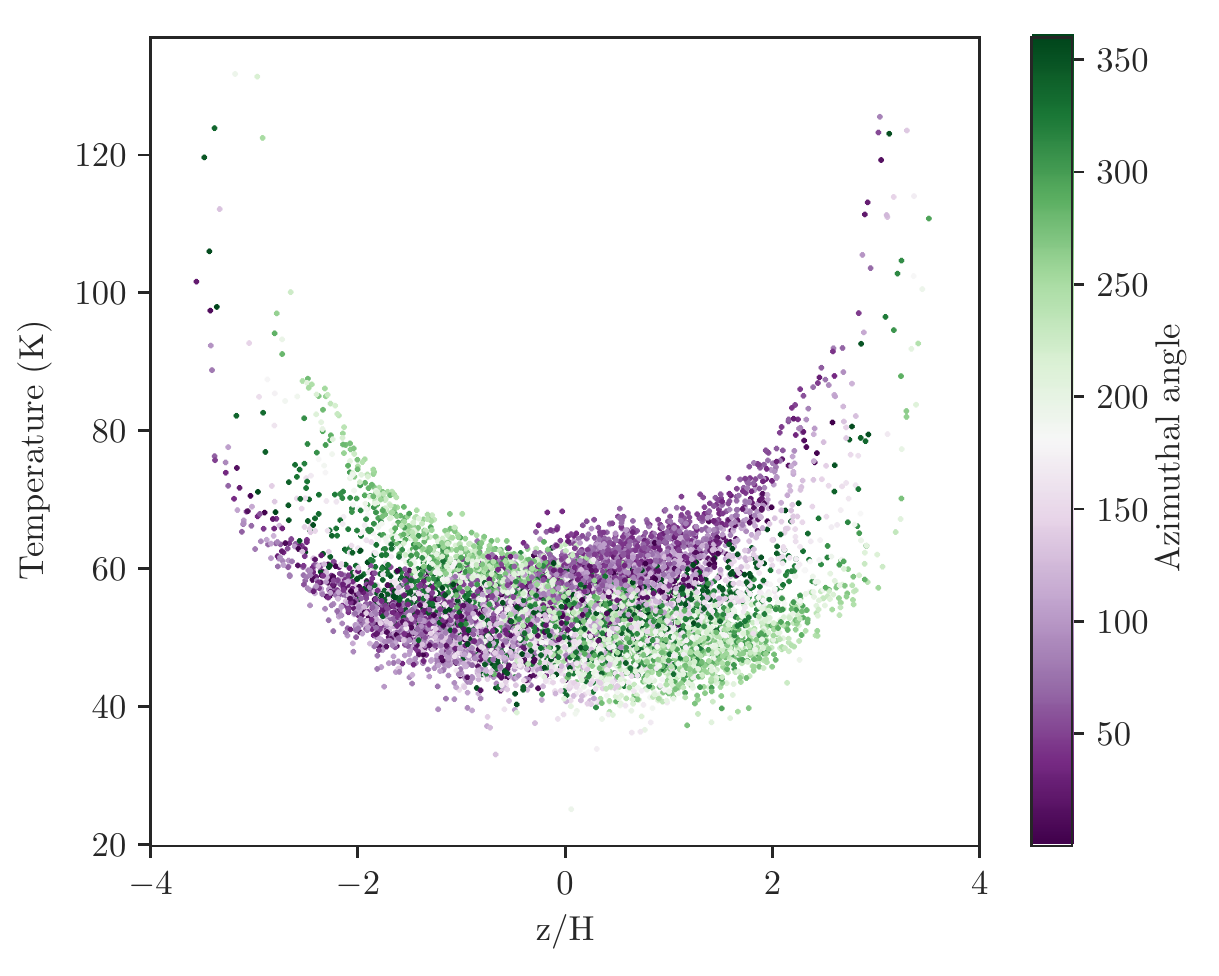}
   \caption{Effect of shadowing on the temperature profile in the outer disc. The temperature of each of the cells between $9.9<R<10.1$~au used in our scattered light calculation is shown, coloured with the azimuthal angle. When the azimuthal angle is $< 180^{\circ}$ ($> 180^{\circ}$) the upper side of the disc is warm (cool) and the lower side is cool (warm). This is consistent with the shadow cast by the misaligned inner disc.} \label{figure:azimuthal_temp}
\end{figure}

Figure~\ref{figure:azimuthal_temp} shows the azimuthal temperature profile between $9.9 < R < 10.1$~au, with the cells coloured by their azimuthal angle (such a narrow radial annulus is only chosen for plotting convenience, we find the same results with larger annuli). Significant departures from the locally isothermal approximation occur at only more than two scale-heights from the mid-plane.

Additionally, the striking pattern on either side of the mid-plane demonstrates the effect that shadowing has on the temperature structure in the disc. The upper and lower surface of the disc have opposing temperatures at the same azimuthal position, such that one side of the disc is warm while the other is cool. This temperature differential is driven by the shadowing due to the misaligned inner disc and will affect the dust thermal emission and disc scale-height of the disc if taken into account during our simulation.

\section{Casting shadows: an application to TW Hya}
\label{section:TW_Hya}
Recent observations of TW Hya have identified an azimuthal surface brightness variation that is proposed to be created by an inclined inner disc driven by a misaligned planet. From observations, this shadow has two defining features; the speed at which it rotates ($22.7^{\circ}$/yr) and the radial location (outside of $\sim$50~au) \citep{Roberge:2005be,Debes:2017fk,Poteet:2018be}. In Section~\ref{section:scattered_light} we have shown that an inner, misaligned disc generated by an inclined planet is capable of casting shadows with similar properties. We thus seek to test our model by applying it to the shadow observed in TW Hya. Indeed, if this model is consistent with the observations it will place constraints on the location and mass of the misaligned planet. Based on the features of the shadow observed in TW Hya, we consider the location, speed, depth of the shadow and secular evolution of the disc in turn.

\subsection{Location of the shadow}
The shadow identified in TW Hya appears only in the outer disc, where $R\gtrsim 53$~au. \citet{Debes:2017fk} suggest that the absence of the shadow at intermediate radii may be caused by a few different mechanisms. First, they suggest the relative geometry between the inner and outer disc may result in a shadow only being cast in the outer disc. With the arrangement proposed by \citet{Debes:2017fk}, the shadow is likely to appear as in the left panels of Figure~\ref{figure:reference_composite} where we find the shadow is weaker in the outer disc. Second, they propose that the inner rim of the outer disc is puffed in such a way as to cast a shadow that masks any azimuthal variation \citep[as in][]{Dullemond:2001bt,Dong:2015nd}. However, the azimuthal temperature profile in Figure~\ref{figure:azimuthal_temp} demonstrates that if there is a misaligned inner disc it will drive an asymmetry in the temperature structure at the inner edge of the outer disc. This suggests that the puffing would also be non-axisymmetric. Third, \citet{Debes:2017fk} suggest that one of the currently observed gaps in the inner region is capable of interfering with the shadow. While shadows from the puffed edges of a planet gap have been considered \citep{Jang-Condell:2008vt,Jang-Condell:2009qb}, our simulations do not include a planet at larger radii and so cannot address this.

Our scattered light image in the right panels of Figure~\ref{figure:reference_composite} suggests an additional option: warping of the outer disc by a misaligned planet produces comparable fluctuations in the azimuthal surface brightness profile. Crucially, the strength of this effect depends on the radius due to the radial dependence of the warp. Figure~\ref{figure:parameter_sweep_outer_shadow} also demonstrates that a planet massive enough to separate an inner disc (in our simulations, $\gtrsim$2M$_{\rm J}$) is also massive enough to warp the outer disc. Thus any planet that can drive an inner, precessing disc \emph{must} cause warping of the outer disc which would affect the profile of the shadow as a function of the radius. In this scenario, a misaligned planet could self consistently explain two of the key features of the shadow in TW Hya.

\subsection{Depth of the shadow}
The depth of the shadow (measured as the amplitude of the variation in the surface brightness profile, $A$ in Equation~\ref{equation:fit}) is likely to have contributions from both the shadow cast by the misaligned inner disc and the warp in the outer disc. How these effects sum depends on their relative phase; as the warp in the outer disc precesses much more slowly than the precessing inner disc. Thus there must exist particular arrangements where they add constructively and the shadow appears deeper than it otherwise would (and vice versa). For the parameters chosen in our simulations (in particular, the location of the planet orbit and the aspect ratio), Figure~\ref{figure:parameter_sweep_outer_shadow} indicates that either mechanism in isolation could produce variations that are consistent in amplitude to TW Hya for planet masses $\sim 6$M$_{\rm J}$ \citep[$\sim20\%$,][]{Debes:2017fk}.

\subsection{Speed of the shadow}
\label{subsection:speed}
The rate of precession of the inner disc depends strongly on the mass of the planet and its orbital inclination to the precessing disc. In the case that the inner disc is precessing differentially relative to the outer disc, we have additional constraints on the relationship between the planet mass, radial extend of the inner disc and the radial extent of the outer disc.

The frequency of precession of a rigidly precessing circumprimary disc in an inclined binary is given by \citet{Larwood:1996fk},
\begin{align}
\omega_\mathrm{prec} = -\frac{3Gm}{4a^3} \frac{\int_{x_{\mathrm{in}}}^{x_{\mathrm{out}}} \Sigma R^3 dR}{\int_{x_{\mathrm{in}}}^{x_{\mathrm{out}}} \Sigma \Omega_k R^3 dR} \cos \beta'.
\label{equation:larwood}
\end{align}

Here $x_{\rm in}$ and $x_{\rm out}$ are the inner and outer edges of the disc respectively, $m$ is the planet mass, $a$ is the semi-major axis of the orbit, $G$ is the gravitational constant and $\beta'$ is the angle between the disc and the inclined planet. This expression can be used to calculate the precession time-scale, as $t_{\rm prec} = 2\pi/ \omega_\mathrm{prec}$. If we assume $\Sigma \propto R^{-1}$ and a very small inner edge, this reduces to the form used by \citet{Debes:2017fk}\footnote{In deriving Equation~\ref{equation:lai} from \citet{Lai:2014ue}, we find a different leading coefficient to that in Equation 2 of \citet{Debes:2017fk} by a factor of $50^{3/2}$.} \citep[originally from ][]{Lai:2014ue},
\begin{align}
P = \frac{7.21 \times 10^7}{\mu_c \cos i_{\rm c}} \left( \frac{M_*}{M_{\odot}} \right)^{-1/2} \left( \frac{r_{\rm disc}}{\text{1 au}} \right)^{-3/2} \left( \frac{a_{\rm c}}{\text{300 au}} \right)^3 \text{yr},
\label{equation:lai}
\end{align}
where $\mu_{\rm c}$ is the mass ratio of the secondary to the primary, $i_{\rm c} \equiv \beta'$ in Equation~\ref{equation:larwood}, $r_{\rm disc} \equiv x_{\rm out}$ and $a_{\rm c}$ is the semi-major axis of the secondary. Following \citet{Debes:2017fk}, we can solve either of these equations for the position and mass of the planet that will drive precession at the rate observed in TW Hya.

The solution to both Equations \ref{equation:larwood} and \ref{equation:lai} for $t_{\rm prec} = P = 15.9$ years is shown in Figure~
 \ref{figure:TW_Hya_planet}. In this representation, a planet must lie in the vicinity of the solid and dashed lines in order to cause precession of a disc that agrees with the rate observed in TW Hya. Here we have assumed the disc is truncated at 0.05~au \citep[although it may be smaller than this,][]{Johnstone:2014rb}, the disc extends to 90\% of the planet's orbit \citep[as in][]{Debes:2017fk}, the star has a mass of $0.7$M$_{\odot}$ \citep{Huelamo:2008be} and the inclination between the disc and the planet is small. Additionally, we assume a $\Sigma$ profile in Equation~\ref{equation:larwood} that has a taper at the inner edge such that $\Sigma (R_{\rm in}) = 0$. Both of these relations agree that in order to drive the observed precession rate, such a planet must be massive ($\gg$ the gap clearing mass) and relatively close in. For example, a planet of $12.5$M$_{\rm J}$ located at $\sim 0.2$~au would drive a disc between 0.05 and 0.2~au to precess with a period of 15.9 years. These planet masses are broadly consistent with previous predictions by \citet{Facchini:2014jg}.

Radial velocity (RV) observations of pre-main-sequence stars are challenging, due to the high levels of stellar activity, but if we treat the observations of \citet{Setiawan:2008ye} and \citet{Huelamo:2008be} as upper limits on the mass of any companion in TW Hya then they yield a combined upper limit to the RV amplitude of 238$\pm$9~m~s$^{-1}$; any putative planet in the system must induce a stellar reflex velocity below this limit. Assuming a viewing inclination of 7$^{\circ}$, we solve for the planet orbit and mass that will cause a radial velocity measurement of 238$\pm$9~m~s$^{-1}$. The relationship between the planet mass and radius is shown in Figure~\ref{figure:TW_Hya_planet} by the dotted line and represents an upper limit for the mass of the planet --- a planet that is more massive than this should have been revealed in existing observations. This comparison suggests that it is possible for a planet that is massive enough to cause a precessing, inner disc to remain undetected by current radial velocity measurements.

\subsection{Long term evolution}
Our simulations in \citetalias{Nealon:2018ic} are limited in that they do not complete a full precession of the inner disc. We instead consider the viscous and alignment time-scales to infer the long term evolution.

The mode of differential precession invoked to cast the moving shadow requires that the planet and inner disc have some mutual misalignment (Equation~\ref{equation:larwood}). This was evident in \citetalias{Nealon:2018ic}, where we see differential precession cease after $t\sim50$ orbits, when the planet and inner disc show the same tilt (Figures 8 and 9 of that paper). When the inner disc and planet are aligned, a shadow will still be cast (the inner disc will still be misaligned with respect to the outer disc) but it will move with the much slower precession rate of the entire disc. The rate at which this alignment occurs between the planet and the inner disc is on the order of the precession time-scale \citep{Scheuer:1996lr} --- this is $\sim$15.9 years in the case of TW Hya. However, this azimuthal feature has been identified in observations spanning about 17 years \citep{Roberge:2005be,Debes:2017fk,Poteet:2018be}. Given the length of time this shadow should be visible compared to the disc lifetime, it is thus unlikely that the inner disc is driven by a single misaligned planet.

\begin{figure}
\includegraphics[width=\columnwidth]{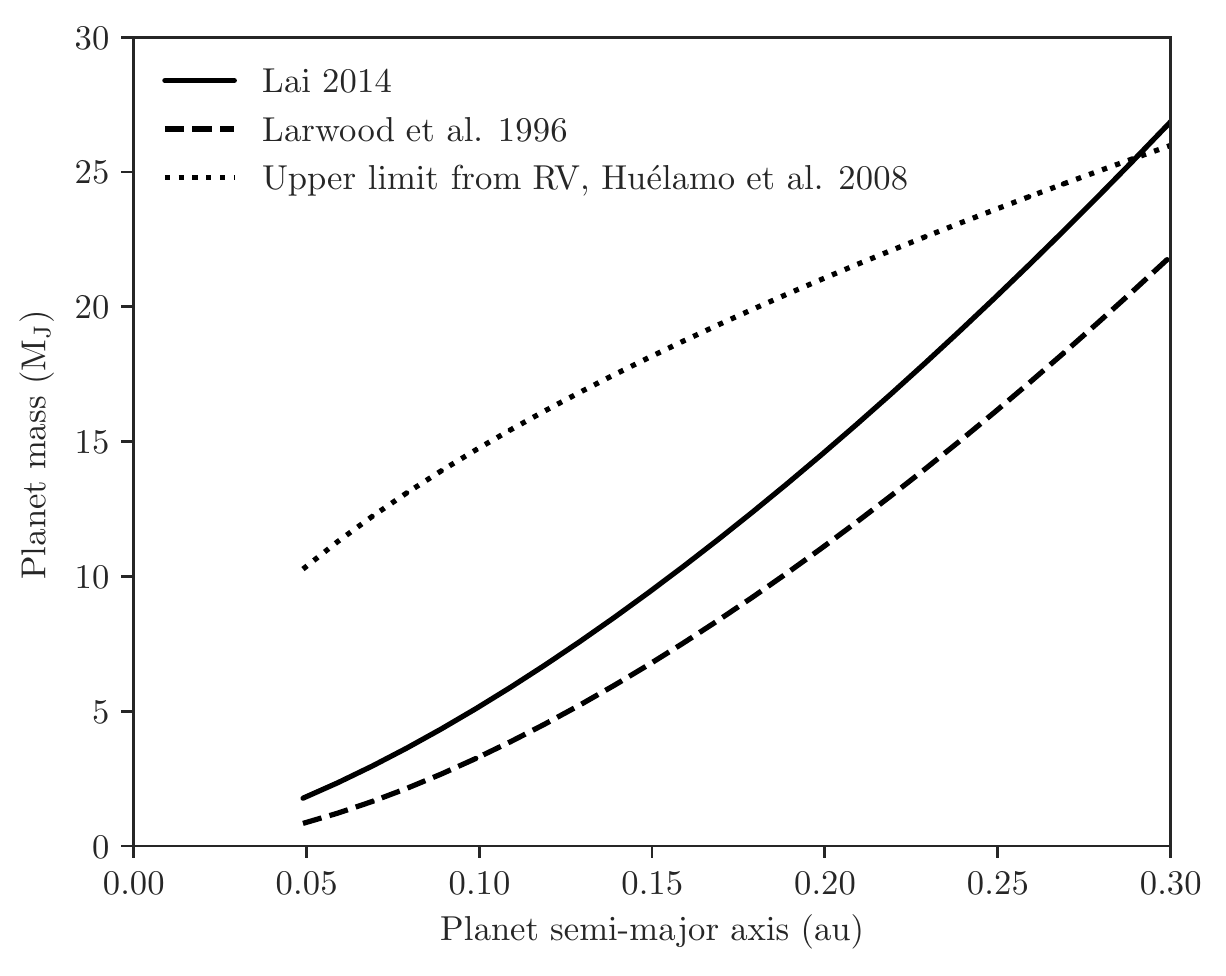}
   \caption{Approximate relationship between mass and radius for a misaligned planet located in TW Hya, based on the observed rate of the shadow in the outer disc (i.e. Equations~\ref{equation:larwood}~and~\ref{equation:lai}). Thus, in order to cast a shadow moving with a precession rate of 15.9 years, the planet would have to be massive and close to the innermost edge of the disc. The innermost edge of the disc fixed at 0.05~au \citep{Eisner:2006uf,Johnstone:2014rb}. The upper limit (dotted line) is calculated for a reflex velocity of $K=238$m/s and $i=7^{\circ}$ \citep{Setiawan:2008ye,Huelamo:2008be}.} \label{figure:TW_Hya_planet}
\end{figure}

Independent of the mechanism that causes the disc to precess, its lifetime may be limited by how quickly it is accreted onto the star. The accretion rate in TW Hya is estimated to be $\sim 1.8 \times 10^{-9}$ M$_{\odot}$/yr \citep{Ingleby:2013ov}. We relate the mass of the disc inside the planet orbit to this accretion rate to find the expected disc life-time, optimistically estimating that the disc could extend to 0.3~au (from Figure~\ref{figure:TW_Hya_planet}). Assuming a surface density profile of \citep{Pringle:1981fo}
\begin{align}
\Sigma(R) = \Sigma_0 \left( \frac{R}{R_0} \right)^{-p} \left(1 - \sqrt{ \frac{R_{\rm in}}{R}} \right),
\end{align}
where $\Sigma_0$ is calibrated to the total disc mass, $R_0=1$~au and $p=1$, the disc mass from the inner edge of the disc $R_{\rm in} = 0.05$~au to the planet orbit at $R_p=0.3$~au is
\begin{align}
M_{\rm disc} = \int_{R_{\rm in}}^{R_{\rm p}}{2 \pi R \Sigma(R)}dR.
\end{align}
Assuming a total disc mass (i.e. between 0.1~au and 50~au) of $5 \times 10^{-3}$ M$_{\odot}$ as in our simulations, this suggests the inner disc life-time of $\sim 6.3 \times 10^3$ years or less than $\sim0.1$\% of the total disc life-time of $\sim10^7$ years \citep[e.g.][]{Weinberger:2013yw}. This measurement is similar to the estimation using the viscous life-time, expressed as
\begin{align}
    t_{\nu} = \left(\frac{H}{R} \right)^{-2} \frac{1}{\alpha \Omega_{\mathrm k}},
\end{align}
where $H/R$ is the aspect ratio, $\alpha$ the Shakura-Sunyaev viscosity parameter and $\Omega_{\rm k}$ the Keplerian angular velocity. Here we set $H/R = 0.05 (R/R_0)^{1/2 - q}$, with $q=1/4$, $R_0=1$~au and $\alpha=10^{-3}$. At $R=0.3$~au the viscous time-scale is then $\sim 2.3 \times 10^4$ years. It is thus highly unlikely that this disc can be sustained long enough to be observed unless it is being fed by the outer disc, but such replenishment is not expected across the gap that must exist in order for the precession of the inner disc to occur.

We note that the above estimate is directly proportional to the total mass of the disc but this is difficult to measure \citep{Thi:2010vy,Bergin:2013ai,Trapman:2017gw,Teague:2018vw}. For the largest realistic total disc mass of $0.1$ M$_{\odot}$ the inner disc life-time increases to $1.2 \times 10^5$ years or $\sim 1.2$\% of the disc life-time. The life-time of this inner, precessing disc is thus much shorter than the estimated disc life-time, even taking into account the uncertainty in the total disc mass.

\section{Discussion}
\label{section:disc}
Observations of a moving azimuthal brightness pattern in TW Hya have been proposed to be caused by a misaligned, precessing inner disc casting a shadow on the outer disc \citep{Debes:2017fk,Poteet:2018be}. Although our results in Section~\ref{section:scattered_light} show that the observational signatures of a disc misaligned by a planet may provide a consistent explanation, our discussion in Section~\ref{section:TW_Hya} shows that this is unlikely to be the complete picture. We thus consider other mechanisms that may result in shadowing from the inner disc and the relevant limitations of our simulations from \citetalias{Nealon:2018ic}.

\subsection{Alternative mechanisms to misalign the inner disc}
While precession of the inner disc due to a single misaligned planet is unlikely, the interaction of multiple bodies may be able to sustain precession. For example, the presence of another massive planet may lead to planet-planet scattering interactions that could continue to perturb the inner planet \citep{Thommes:2003is}. Indeed, there is circumstantial evidence of planets outside of 1~au in TW Hya \citep{Calvet:2002vr,Debes:2013yj,Andrews:2016bw,Tsukagoshi:2016hf,vanBoekel:2017ce,Mentiplay:2018vr}. Additionally, if the outer disc is particularly massive it may affect the precession rate of the inner disc. While our simulations only consider a disc mass of $5\times10^{-3}$~M$_{\odot}$, the observed disc mass is poorly constrained \citep{Thi:2010vy,Bergin:2013ai,Trapman:2017gw,Teague:2018vw} and may be massive enough to influence the inner disc precession.

One limitation of our simulations in \citetalias{Nealon:2018ic} is that they do not take into account any magnetic field effects. At the predicted location of the misaligned planet (Figure~\ref{figure:TW_Hya_planet}) the stellar magnetic field is likely to interact with the disc. An example of how this may affect the inner edge of the disc is found in AA Tau, which has a stellar magnetic field that is misaligned by about 20$^{\circ}$ \citep{Bouvier:1999uh,Donati:2010be}. Non-ideal MHD simulations by \citet{Romanov:2004br} demonstrated that the inner edge of a disc warps under the influence of the misaligned magnetic field. This type of warping represents an additional method to cast a shadow from the inner disc to the outer disc, but here the time-scale for the shadow to move would be related to the stellar magnetic field structure. In applying such a model to TW Hya, we note that measurements of the magnetic field geometry suggest it is misaligned by $\lesssim 10^{\circ}$ \citep{Donati:2011bt} and the inner radius of the disc is 0.05~au \citep{Eisner:2006uf,Johnstone:2014rb}. Thus if the shadow cast from the misaligned inner disc was due to magnetic fields it would likely have a similar time-scale to the stellar rotation period, which is observed to be days in TW Hya. It is not clear how time-scales of years required by the observed shadow in TW Hya can be related to the much shorter time-scales due to stellar magnetic field effects at the inner edge of the disc.

Our simulations are also limited by their radial extent and duration. In \citetalias{Nealon:2018ic} we demonstrated that the misalignment of the inner disc and the warp structure in the outer disc both evolve dramatically across the 200 planet orbits simulated. Both the variation due to the misaligned inner disc and the warped outer disc decrease across the last 100 orbits of our simulations. We thus caution that the amplitudes presented in Figures~\ref{figure:reference_composite}~and~\ref{figure:parameter_sweep_outer_shadow} are optimistic, but note that the other features of the shadow (i.e. radial location) do not change appreciably for $t>100$ orbits. The radial profile of the warp under the influence of multiple planets also remains unexplored.

\subsection{Alternative mechanisms to cast a shadow}
The shadow in TW Hya may alternatively be the cast by the dense tidal stream when the planet is on the near side of the disc. We find hints of this in the earlier stages of our simulations, when the planet is still rapidly accreting. In this case the speed of the shadow may be explained by the Keplerian orbit of a misaligned planet at 5.6~au. Additionally, it implies that the shadow would only be observed for the half of the orbit when the planet is on the near side of the disc. This is consistent with current observations (e.g. Figure 7 of \citealt{Debes:2017fk}) and suggests that the shadow will not be visible after $\sim$2022.

The radial dependence of the shadow observed in TW Hya may be self-consistently explained if the disc is warped by the presence of a misaligned planet. However, warping of the outer disc could also be caused by a smaller, misaligned planet at larger radii or by a previous interaction with an external perturber. Recent simulations by \citet{Cuello:2018bd} showed that the disc may be warped by such an episode and retain the warp long after the interaction \citep[see also][]{Nixon:2010iq}. For the relatively isolated TW Hya, such an interaction appears unlikely. In any case, we note that the time-scale of the shadow cannot be matched exclusively by a warp in the outer disc.

\section{Conclusions}
\label{section:concs}
We have examined the signatures in scattered light that are caused in a protoplanetary disc under the influence of a misaligned planet. Using our simulations from \citetalias{Nealon:2018ic}, we have conducted radiative transfer calculations using {\sc mcfost}. We have then examined the azimuthal profile of the surface brightness from both a shadow cast from an inner, misaligned disc and the variation due to a warp in the outer disc. Our results are summarised as:
\begin{enumerate}
\item An inner disc misaligned by more than a degree can cast non-axisymmetric shadows on the outer disc which are observable in IR scattered light (with a shadow depth of $\gtrsim 20\%$ compared to the average flux at that radius). This shadow is strongest closest to the inner edge of the outer disc and decreases in intensity at larger radii. The azimuthal location of this shadow is consistent with a shadow cast from the inner disc. In the case that the disc is created by a misaligned planet, our simulations suggest the planet must be $\gtrsim6.5$M$_{\rm J}$ and inclined by more than $\sim2^{\circ}$ to cast an appreciable shadow. These values are specific to our model, but correspond to a planet with 5.0$\times$ the thermal masses at an inclination of 0.5$\times H/R$. In contrast to previous works, a only small relative misalignment between the inner and outer disc is required to produce shadowing (and this inclination is less than the initial inclination of the planet).
\item The presence of a warp can additionally cause measurable variations in the azimuthal surface brightness. The amplitude of this variation depends on the relative warp between the star and the disc, and thus can change the appearance of this variation as a function of radius. In our simulations with a misaligned planet, the variation is strongest at large radii. Our simulations suggest that planets that are massive enough to cause a misaligned inner disc also drive a warp in the outer disc.
\item Our simulations suggest that a misaligned inner disc may be consistent with observations of the moving shadow in TW Hya. Despite this, we argue that it is difficult to sustain this configuration over observationally relevant time-scales, making this an unlikely arrangement for TW Hya.
\end{enumerate}

Although the scattered light images from a misaligned inner disc are consistent with observations, we suggest that a single misaligned planet is unlikely to be the mechanism driving such a disc. Alternative scenarios include a misaligned planet in the outer disc providing the outer disc warp, the inner disc warped by a misaligned stellar magnetic field, Keplerian motion of a misaligned inner planet at $5$~au or precession driven by more than a single planet.

Even the case that the misaligned disc is not driven by a planet, this work demonstrates that small to moderate misalignments in the disc are capable of casting shadows onto the outer disc. In combination with previous studies looking at strongly misaligned discs \citep{Facchini:2017of,Zhu:2018vf}, this confirms that warps and misalignments can be used to inform on the underlying dynamics in protoplanetary discs.

\section*{Acknowledgements}
The authors warmly thank Daniel Price, Cassandra Hall, Chris Nixon and Jim Pringle for useful discussions. We also thank John Debes and the referee for constructive comments. This project has received funding from the European Research Council (ERC) under the European Union's Horizon 2020 research and innovation programme (grant agreement No 681601). DM is funded by a Research Training Program Stipend from the Australian government. This work was performed using the DiRAC Data Intensive service at Leicester, operated by the University of Leicester IT Services, which forms part of the STFC DiRAC HPC Facility (www.dirac.ac.uk). The equipment was funded by BEIS capital funding via STFC capital grants ST/K000373/1 and ST/R002363/1 and STFC DiRAC Operations grant ST/R001014/1. DiRAC is part of the National e-Infrastructure. Figure~\ref{figure:density_cut} and the upper panels of Figure~\ref{figure:reference_composite} were plotted with {\sc splash} \citep{Price:2007kx}. All other figures were produced using the community open-source Python package Matplotlib \citep{matplotlib}.




\bibliographystyle{mnras}
\bibliography{master} 
\label{lastpage}
\end{document}